\documentclass[sigconf]{acmart}
\usepackage{caption}
\usepackage{subcaption}
\usepackage{enumitem}
\usepackage{mdwlist}
\usepackage{footnote}
\usepackage{multirow}
\usepackage[ruled,linesnumbered]{algorithm2e}
\setlist{nosep}

\AtBeginDocument{%
  \providecommand\BibTeX{{%
    \normalfont B\kern-0.5em{\scshape i\kern-0.25em b}\kern-0.8em\TeX}}}

\copyrightyear{2020}
\acmYear{2020}
\setcopyright{acmcopyright}\acmConference[SIGIR '20]{Proceedings of the 43rd International ACM SIGIR Conference on Research and Development in Information Retrieval}{July 25--30, 2020}{Virtual Event, China}
\acmBooktitle{Proceedings of the 43rd International ACM SIGIR Conference on Research and Development in Information Retrieval (SIGIR '20), July 25--30, 2020, Virtual Event, China}
\acmPrice{15.00}
\acmDOI{10.1145/3397271.3401174}
\acmISBN{978-1-4503-8016-4/20/07}

\begin{document}
\fancyhead{}	

\title{Interactive Recommender System via Knowledge Graph-enhanced Reinforcement Learning}
	
\author{Sijin Zhou$^1$, Xinyi Dai$^1$, Haokun Chen$^1$, Weinan Zhang$^1$, Kan Ren$^1$}
\author{Ruiming Tang$^2$, Xiuqiang He$^2$, Yong Yu$^1$}
% \affiliation{$^1$Shanghai Jiao Tong University, $^2$Huawei Noah's Ark Lab}
% \email{{zhousijin, xydai, chenhaokun, wnzhang, kren, yyu}@apex.sjtu.edu.cn,  {tangruiming, hexiuqiang1}@huawei.com}

\affiliation{
 \institution {$^1$Shanghai Jiao Tong University, $^2$Huawei Noah's Ark Lab \\
 \{zhousijin, xydai, chenhaokun, wnzhang, kren, yyu\}@apex.sjtu.edu.cn, \{tangruiming, hexiuqiang1\}@huawei.com\\~}
}

\begin{abstract}
Interactive recommender system (IRS) has drawn huge attention because of its flexible recommendation strategy and the consideration of optimal long-term user experiences. 
To deal with the dynamic user preference and optimize accumulative utilities, researchers have introduced reinforcement learning (RL) into IRS. 
However, RL methods share a common issue of sample efficiency, i.e., huge amount of interaction data is required to train an effective recommendation policy, which is caused by the sparse user responses and the large action space consisting of a large number of candidate items.
Moreover, it is infeasible to collect much data with explorative policies in online environments, which will probably harm user experience.
In this work, we investigate the potential of leveraging knowledge graph (KG) in dealing with these issues of RL methods for IRS, which provides rich side information for recommendation decision making.
Instead of learning RL policies from scratch, we make use of the prior knowledge of the item correlation learned from KG to (i) guide the candidate selection for better candidate item retrieval, (ii) enrich the representation of items and user states, and (iii) propagate user preferences among the correlated items over KG to deal with the sparsity of user feedback. Comprehensive experiments have been conducted on two real-world datasets, which demonstrate the superiority of our approach with significant improvements against state-of-the-arts.
\end{abstract}


\settopmatter{printacmref=false, printfolios=false}

\maketitle	

{\fontsize{8pt}{8pt} \selectfont
\textbf{ACM Reference Format:}\\
Sijin Zhou, Xinyi Dai, Haokun Chen, Weinan Zhang, Kan Ren, Ruiming Tang, Xiuqiang He and Yong Yu. 2020. Interactive Recommender System via Knowledge Graph-enhanced Reinforcement Learning. In \textit{ Proceedings of the 43rd International ACM SIGIR Conference on Research and Development in Information Retrieval (SIGIR'20), July 25--30, 2020, Virtual Event, China.} ACM, NY, NY, USA, 10 pages. https://doi.org/10.1145/3397271.3401174 }

\section{Introduction}\label{sec:intro}

	With the wide use of mobile applications such as TikTok, Pandora radio and Instagram feeds, interactive recommender systems (IRS) have received much attention in recent years \cite{zhao2013interactive,wang2017factorization}. 
	Unlike that in traditional recommender systems \cite{he2017neural,wang2006unifying, koren2009matrix}, where the recommendation is treated as a one-step prediction task, the recommendation in IRS is formulated as a multi-step decision-making process. In each step, the system delivers an item to the user and may receive feedback from her, which subsequently derives the next recommendation decision in a sequential manner. The recommendation-feedback interaction is repeated until the end of this visit session of the user. The goal of the IRS is to explore users' new interests, as well as to exploit the learned preferences, to provide accurate predictions, so as to optimize the outcome of the entire recommendation sequence \cite{zhao2013interactive,zhao2018recommendations}.
	
	One way to implement IRS and balance the exploration and exploitation is multi-armed bandit (MAB) methods~\cite{li2010contextual,zeng2016online,wang2017factorization}. In MAB-based models, the user preference is often modeled by a linear function that is continuously learned through the interactions with proper exploration-exploitation tradeoff.
	However, these MAB-based models pre-assume that the underlying user preference remains unchanged during the recommendation process, i.e., they do not model the dynamic transitions of user preferences \cite{zhao2013interactive}.
	The key advantage for modern IRS is to learn about the possible dynamic transitions of the user's preference and optimize the long-term utility.
	
	Recently, some researchers have incorporated deep reinforcement learning (DRL) models into interactive recommender system \cite{zheng2018drn,zhao2018recommendations,hu2018reinforcement,chen2019large, zhao2018deep}, due to the great potential of DRL in decision making and long-term planning in dynamic environment~\cite{silver2016mastering}.  \citet{mahmood2007learning} first proposed to used model-based techniques in RL where dynamic programming algorithms such as policy iteration are utilized. Some recent works use model-free frameworks to tackle IRS tasks, e.g., deep Q-network (DQN)~\cite{zhao2018recommendations} and deep deterministic policy gradient (DDPG)~\cite{hu2018reinforcement}.

	Nevertheless, employing DRL in real-world interactive recommender system is still challenging.
	A common method for training recommendation models is to make use of offline logged data directly, but it will suffer from the estimation bias problem~\cite{chen2019top} under the real-time interaction setting. In contrast, the ideal setting of learning the optimal recommendation policy is to train the agent online. However, due to the item-state search space for each recommendation step and the trial-and-error nature of RL algorithms, 
	DRL methods normally face sample efficiency problem \cite{zou2019reinforcement}, i.e., learning such a policy requires a huge amount of data through interacting with real users before achieving the best policy, which may degrade user experience and damage system profit \cite{zhang2016collectivepolicy}. Therefore, it is quite crucial to improve the sample efficiency of existing DRL models with only a limited amount of interaction data.
	
	Fortunately, there is rich prior knowledge from other external sources that may contribute to dealing with the above problems, such as textual reviews, visual images or item attributes~\cite{cheng2016wide}.
	Among these, knowledge graph (KG), a well-known structured knowledge base,
	represents various relations as the attributes of items and links items if they have common attributes, which has shown great effectiveness for representing the correlation between items \cite{wang2018ripplenet}.  The association between the items provided by KG is very suitable for the recommendation scenarios. For example, a user likes the movie $\textit{Inception}$, and the information behind it may be that her favorite director is $\textit{Nolan}$.  With such links among actions (items) on the graph, one user-item interaction record could reveal the user's preference on multiple connected items. In addition, the information contained in the semantic space of the entire knowledge graph will also be helpful in extracting user interest during the recommendation process. Since there have been many successful works applying open-sourced KGs (such as DBpedia, NELL, and Microsoft Satori) %DBpedia\footnote{https://wiki.dbpedia.org/}, NELL\footnote{http://rtw.ml.cmu.edu/rtw/} and Microsoft Satori\footnote{https://searchengineland.com/library/bing/bing-satori}) 
	to \textit{traditional} recommendation systems~\cite{yu2014personalized,zhang2016collaborative,wang2018ripplenet}, we believe that it is reasonably promising to leverage KG to DRL-based methods in IRS scenarios.

	In this paper, we make the first attempt to leverage KG for reinforcement learning in interactive recommender systems, trying to address the aforementioned limitations of the existing DRL methods.
	We propose \textit{KGQR (Knowledge Graph enhanced Q-learning framework for interactive Recommendation)}, a novel architecture that extends DQN. Specifically, we integrate graph learning and sequential decision making as a whole to facilitate knowledge in KG and pattern mining in IRS. On one hand, to alleviate data sparsity, the user feedback is modeled to propagate via structure information of KG, so that the user's preference can be transited among correlated items (in the KG).  In this way, one interactive record can affect multiple connected items, thus the sample efficiency is improved. On the other hand, by aggregating the semantic correlations among items in KG, the item embedding and the user's preference are effectively represented, which leads to more accurate Q-value approximation and hence better recommendation performance. In addition, we also conduct a methodology to deal with action selection in such large space. Rather than enumerating the whole item set, each step the candidate set for recommendation is dynamically generated from the local graph of KG, by considering the neighborhood of the items in user's high-scored interacted items. The method of candidate selection forces the deep Q-network to fit on the samples that KG considers more useful through the structure of item correlations, hence it can make better use of limited learning samples for RL-algorithm.

	To the best of our knowledge, this is the first work to introduce KG into RL-based methods to interactive recommender systems. The contributions of our work can be summarized as follows.
	\begin{itemize}[leftmargin = 10pt]
		\item We propose a novel end-to-end deep reinforcement learning based framework KGQR for interactive recommendation to addresses the sparsity issue. By leveraging prior knowledge in KG in both candidate selection and the learning of user preference from sparse user feedback, KGQR can improve sample efficiency of RL-based IRS models.
		\item The dynamic user preference can be represented more precisely by considering the semantic correlations of items in KG, with graph neural networks.
		\item Extensive experiments have been conducted on two real-world datasets, demonstrating that KGQR is able to achieve better performance than state-of-the-arts with much fewer user-item interactions, which indicates high sample efficiency.
	\end{itemize}

	\section{Related Work}
	\label{sec:related}
	\noindent\textbf{Traditional KG Enhanced Recommendation}.
	Traditional KG enhanced recommendation models can be classified into three categories: path-based methods, embedding-based methods and hybrid methods. 
	In \emph{path-based} methods~\cite{yu2014personalized,shi2015semantic,zhao2017meta}, KG is often treated as a heterogeneous information network (HIN), in which specific meta-paths/meta-graphs are manually designed to represent different patterns of connections.
	The performance of these methods is heavily dependent on the hand-crafted meta-paths, which are hard to design.
	In \emph{embedding-based} methods, the entity embedding extracted from KG via Knowledge Graph Embedding (KGE) algorithms (like TransE~\cite{bordes2013translating}, TransD~\cite{ji2015knowledge}, TransR~\cite{lin2015learning}), is utilized to better represent items in recommendation. \citet{zhang2016collaborative} propose Collaborative Knowledge Base Embedding (CKE), to jointly learn the latent representations in collaborative filtering as well as items' semantic representations from the knowledge base, including KG, texts, and images. MKR~\cite{wang2019multi} associates the embedding learning on KG with the recommendation task by cross \& compress units. 
	KSR~\cite{huang2018improving} extends the GRU-based sequential recommender by integrating it with a knowledge-enhanced Key-Value Memory Network.
	In \emph{hybrid} methods, researchers combine the above two categories to learn the user/item embeddings by exploiting high-order information in KG. Ripplenet~\cite{wang2018ripplenet} is a memory-network-like model that propagates users' potential preferences along with links in the KG.
	Inspired by the development of graph neural network ~\cite{kipf2016semi,hamilton2017inductive,velivckovic2017graph}, KGAT ~\cite{wang2019kgat} applies graph attention network~\cite{velivckovic2017graph} framework in a collaborative knowledge graph to learn the user, item and entity embeddings in an end-to-end manner.

	However, most of these methods are one-step prediction tasks and can not model the iterative interactions with users. Besides, they all greedily optimize an immediate user's feedback and
	don't take the user's long-term utility into consideration.

	\vspace{5pt}\noindent\textbf{Reinforcement Learning in IRS}.
	RL-based recommendation methods model the interactive recommendation process as a Markov Decision Process (MDP), which can be divided into model-based and model-free methods. As one of the \emph{model-based} techniques, Mahmood and Ricci \cite{mahmood2007learning} utilize policy iteration to search for the optimal recommendation policy where an action is defined to be an item and a state is represented as $n$-gram of items. 
	Policy iteration needs to go through the whole state space in each iteration, with exponential complexity to the number of items. Therefore, it is unable to handle large state and action space.
	
	Recently, most works on RL-based recommendation prefer \emph{model-free} techniques, including policy gradient (PG)-based, DQN-based and DDPG-based methods. PG-based methods, as well as DQN-based, treat recommending an item as an action. PG-based methods~\cite{chen2019large} learn a stochastic policy as a distribution over the whole item space and sample an item according to such distribution. 

	DQN-based methods ~\cite{zhao2018recommendations,zheng2018drn,zou2019reinforcement} learn Q-value for each item and select the item with the maximum Q-value.
	Zheng et al.~\cite{zheng2018drn} combine DQN and Dueling Bandit Gradient Decent (DBGD) \cite{grotov2016online} to conduct online news recommendation. Zou et al.~\cite{zou2019reinforcement} integrate both the intent engagement (such as click and order) and long-term engagement (such as dwell time and revisit) when modeling versatile user behaviors. In DDPG-based works, an action is often defined as a continuous ranking vector. 
	Dulac et al.~\cite{dulac2015deep} represent the discrete actions in a continuous vector space, pick a proto-action in a continuous hidden space according to the policy and then choose the valid item via a nearest neighbor method. In \cite{hu2018reinforcement, zhao2018deep}, the policies compute the ranking score of an item by calculating the pre-defined function value (such as an inner product) of the generated action vector and the item embedding.
	
	Nevertheless, all existing RL-based recommendation  models suffer from low sample efficiency issue and need to pre-train user/item embeddings from history, which means that they cannot handle recommendation on cold-start problem well. A significant difference between our approach and the existing models is that we first propose a framework that combines the semantic and structural information of KG with the IRS to break such limitations.

	\begin{figure}[t]
		\begin{centering}
			\includegraphics[width=\columnwidth]{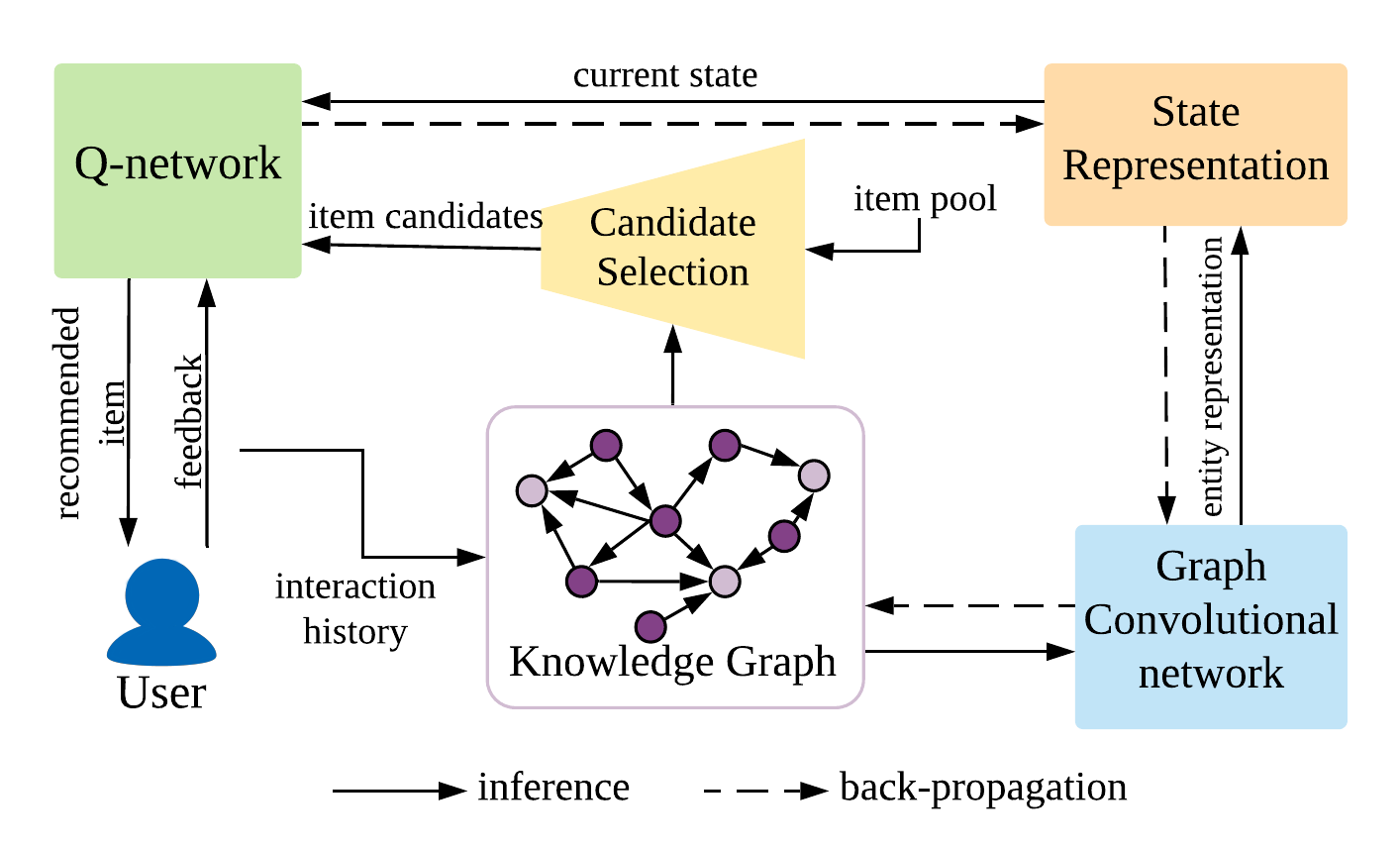}
			\vspace{-10pt}
			\caption{The overall KGQR framework. The left part illustrates the interaction process between the user and the IRS. The right part illustrates how the IRS recommends an item to the user according to past interaction history.}
			\label{fig:overall} 
		\end{centering}
		\vspace{-10pt}
	\end{figure}

	\section{Problem Formulation}
	\label{sec:problem}
	In feed streaming recommendation scenario, the interactive nature between the recommender system and the user is a multi-step interaction process that lasts for a period of time. At each timestep $t$, according to the observations on past interactions,
	the recommendation agent delivers an item $i_t\in\mathcal I$ to the user, and receives feedback (e.g., click, purchase or skip) from her. This process continues until the user leaves the recommender system. Under such circumstances, the interactive recommendation process can be formulated as a Markov Decision Process (MDP). The ultimate goal of the recommender system is to learn a recommendation policy $\pi:\mathcal S \to \mathcal I$, which maximizes the cumulative utility over the whole interactive recommendation process as 
	\begin{equation}
	\pi^* = \arg\max_{\pi\in\Pi}\mathbb{E}\Big[\sum_{t=0}^{T} r(s_t,i_t)\Big] .
	\end{equation}
	Here $s_{t}\in \mathcal{S}$ is a representation abstracted from user's positively interacted items $ o_t = \{i_1,i_2,...i_n\}$ that denotes user's preference at timestep $t$; $r(s_t,i_t)$ is the user's immediate feedback to the recommended item $i_t$ at the state $s_t$ according to some internal function $\mathcal{R} : \mathcal{S} \times \mathcal{I} \to \mathbb{R}$, abbreviated as $r_t$.

	\begin{table}[t]
		\centering
		\caption{Notations and descriptions.}
		\vspace{-10pt}
		\resizebox{\columnwidth}{!}{
			\setlength{\tabcolsep}{3pt}
			\begin{tabular}{c|l}
				\hline
				Notations & Descriptions \\
				\hline
				$\mathcal U, \mathcal I$ & Set of users and items in IRS.\\
				$\mathcal G = (\mathcal E, \mathcal R)$ & Knowledge graph. \\
				$\mathcal E, \mathcal R $ & Set of entities and relations in $\mathcal G$. \\
				$ o_t = \{i_1,i_2,...i_n\}$ & Recorded user's positively interacted at timestep $t$. \\
				$\mathbf{s}_t $ & Dense representation of user's preference at timestep $t$. \\
				$ r_t $ & The user's reward at timestep $t$. \\
				$T$ & Episode length. \\
				$\mathbf{e}_{h},h\in \mathcal{E} $ & Dense representation of an entity.\\
				$\mathcal I_t(\mathcal G)$ & Candidate action space at timestep $t$.\\
				$\theta_S$ & Parameters of state representation network.\\
				$\theta_Q,\theta_Q'$ & Parameters of online Q-network /target Q-network.\\
				$\mathcal D$ & Replay Buffer.\\
				\hline
			\end{tabular}
		}
		\label{table:notation}
	\end{table}		
	
	To achieve this goal, traditional recommendation methods usually adopt a greedy strategy and only optimize one-step reward, i.e., at each timestep $t$, they optimize the immediate reward $r_t$. Different from them, 
	DRL algorithms take the long-term impact into consideration and explicitly model the long-run performance. They will optimize $\sum_{j=0}^{T-t} \gamma^{j} r_{t+j}$ at timestep $t$, instead. And $\gamma \in (0, 1] $ is the discount factor to control the degressively accumulated long-term rewards.

	In general, we can use Q-value to evaluate the value of an action (i.e., recommending an item) taken at a given state, defined as
	\begin{equation}
	Q^{\pi}(s_{t},i_{t}) = \mathbb{E}\Big[\sum_{j=0}^{T-t} \gamma^{j} r_{t+j}\Big]=\mathbb{E}\Big[r_{t}+\sum_{j=1}^{T-t}\gamma^{j}r_{t+j} \Big],
	\end{equation}
	which is a weighted sum of the expected reward of all future steps starting from the current state and following the policy $\pi$ to take actions. Then following the optimal Bellman equation~\cite{bellman1952theory}, the optimal $Q^*$, having the maximum expected reward achievable is:  
	\begin{equation}
	Q^{*}(s_{t},i_{t}) = \mathbb{E}_{s_{t+1}}[r_{t}+\gamma \max_{i_{t+1}}Q^{*}(s_{t+1},i_{t+1})|s_{t},i_{t}].
	\label{eq:bellman}
	\end{equation}
	Since the state and action spaces are usually enormous, we normally estimate the Q-value of each state-action ($s_t,i_t$) pair via a $\theta_Q$-parameterized deep neural network, i.e., $Q^*_{\theta_Q}(s_{t},i_{t}) \approx Q^*(s_t,i_t)$. 
	
	As mentioned in Section~\ref{sec:intro}, learning this Q-function from scratch requires numerous interactions with real users due to the low data efficiency problem that is common in the RL algorithm. However, unlike basic RL algorithms, in RS scenarios, KG can provide complementary and distinguishable information for each item by their latent knowledge-level connection in the graph. Thus, with the prior knowledge of the environment and actions, the Q-function can be learned more efficiently, 
	\begin{equation}
	Q^*_{\theta_Q}(s_{t},i_{t};\mathcal G) = Q^*_{\theta_Q}\Big(s_t(\mathcal G),i_t(\mathcal G)\Big).
	\end{equation}
	Here $\mathcal { G } $ is the knowledge graph comprised of subject-property-object triples facts, e.g., triple ($\textit{Nolan}, \textit{DirectorOf}, \textit{Inception}$), which denotes $\textit{Nolan}$ is the director of $\textit{Inception}$. And it is ofen present as ($\textit{head},\textit{relation}, \textit{tail}$), $\textit{head} \in \mathcal{E}$, $ \textit{relation} \in  \mathcal{R}$ , $\textit{tail} \in \mathcal{E}$ and $\mathcal{E}$, $\mathcal{R}$ denote the set of entities and relationships in $\mathcal{G}$, respectively. Usually, an item $i \in \mathcal I$ can be linked to an entity $e \in \mathcal E$ in the knowledge graph, e.g., the movie $\textit{Godfather}$ from MovieLens dataset has a corresponding entity entry in DBpedia. We will introduce how we design the knowledge enhanced DRL framework for IRS in the following sections and the key notations used in this paper are summarized in Table \ref{table:notation}.
	
% 	\begin{table}[t]
% 		\centering
% 		\caption{Notations and descriptions.}
% 		\vspace{-10pt}
% 		\resizebox{\columnwidth}{!}{
% 			\setlength{\tabcolsep}{3pt}
% 			\begin{tabular}{c|l}
% 				\hline
% 				Notations & Descriptions \\
% 				\hline
% 				$\mathcal U, \mathcal I$ & Set of users and items in IRS.\\
% 				$\mathcal G = (\mathcal E, \mathcal R)$ & Knowledge graph. \\
% 				$\mathcal E, \mathcal R $ & Set of entities and relations in $\mathcal G$. \\
% 				$ o_t = \{i_1,i_2,...i_n\}$ & Recorded user's positively interacted at timestep $t$. \\
% 				$\mathbf{s}_t $ & Dense representation of user's preference at timestep $t$. \\
% 				$ r_t $ & The user's reward at timestep $t$. \\
% 				$T$ & Episode length. \\
% 				$\mathbf{e}_{h},h\in \mathcal{E} $ & Dense representation of an entity.\\
% 				$\mathcal I_t(\mathcal G)$ & Candidate action space at timestep $t$.\\
% 				$\theta_S$ & Parameters of state representation network.\\
% 				$\theta_Q,\theta_Q'$ & Parameters of online Q-network /target Q-network.\\
% 				$\mathcal D$ & Replay Buffer.\\
% 				\hline
% 			\end{tabular}
% 		}
% 		\label{table:notation}
% 	\end{table}	

	\section{KGQR Methodology}
	\label{sec:model}

	The overview of our proposed framework is shown in Figure~\ref{fig:overall}. Generally, our KGQR model contains four main components: graph convolution module, state representation module, candidate selection module and  Q-learning network module.  In the interactive recommendation process, at each timestep $t$, the IRS sequentially recommends items $i_t$ to users, and correspondingly updates its subsequent recommendation strategy based on user's feedback $r_t$. At the specific time during one recommendation session, according to the interaction history $o_t$ combined with the knowledge graph $\mathcal G$, the IRS models the user's preference $s_t$ via graph convolution module and state representation module. The details of these two representation learning modules will be discussed in Section~\ref{sec:state}. 
	Then the IRS calculates the highest-scored item in the candidate set through Q-network and recommends it to the user. We will introduce the candidate selection module and deep Q-network module in Section~\ref{sec:CS} and Section~\ref{sec:DDQN}, respectively.

	\subsection{KG Enhanced State Representation}\label{sec:state}
	
	In IRS scenario,  it is impossible to get user's state $s_t$ directly, and what we can directly observe is the recorded user-system interaction history $o_t$. As state is one of the key part in MDP, the design of state representation module is critical to study the optimal recommendation strategy. 
	
	\subsubsection{Graph convolutional embedding layer}
	Generally, the state representation in IRS is abstracted from the user's clicked\footnote{Without loss of generality, we take ``click'' behavior as user positive feedback as the running example.} items, since the positive items represent the key information about what the user prefers~\cite{zhao2018recommendations}. Given the user's history, we first convert the clicked item set $\{i_t\}$ into embedding vectors $\mathbf{i}_t\in \mathbb{R}^{d}$, where $d$ is the dimension of the embeddings. 
	Since we have already linked items with entities in KG, we can take advantage of the semantic and correlation information among items in KG for better item embedding $\mathbf{i}_t(\mathcal G)$. 
	
	In order to distill structural and semantic knowledge in the graph into a low-dimensional dense node representation, different approaches of graph embedding methods can be applied. In addition to harvesting the semantic information, we incline to explicitly link these items so that one data can affect more items. Thus, a graph convolutional network (GCN) ~\cite{kipf2016semi} is used in our work to recursively propagate embeddings along the  connectivity of items and learn the embeddings of all entities $\{\mathbf{e}_{h}\in \mathbb{R}^{d}\}_{h\in \mathcal{E}}$ on the graph $\mathcal G$.

	The computation of the node's representation in a single graph convolutional embedding layer is a two-step procedure: aggregation and integration. These two procedures can naturally be extended to multiple hops, and we use the notation $k$ to identify $k$-th hop.	
	In each layer, first, we aggregate the representations of the neighboring nodes of a given node $h$:
	\begin{equation}
	\mathbf{e}_{N(h)}^{k-1} =  \frac{1}{|N(h)|} \sum_{t \in N(h)} \mathbf{e}_{t}^{k-1},
	\label{eq:GCN1}
	\end{equation}
	where $N(h)=N(\textit{head})=\{\textit{tail} ~|~ (\textit{head},\textit{relation}, \textit{tail} ) \in \mathcal { G }\}$ is the set of neighboring nodes of $h$. Notice that, here we consider the classic \emph{Average} aggregator for example, other aggregator like concat aggregator ~\cite{hamilton2017inductive}, neighbor aggregator or attention mechanism (GAT)~\cite{velivckovic2017graph} can also be implemented.
	
	Second, we integrate the neighborhood representation with $h$'s representation as
	\begin{equation}
	\mathbf{e}_h^{k} = \sigma (\mathbf{W}_{k} \mathbf{e}_{N(h)}^{k-1} +\mathbf{B}_{k} \mathbf{e}_{h}^{k-1}),
	\label{eq:GCN2}
	\end{equation}
	where $\mathbf{W}_{k}$ and $\mathbf{B}_{k}$ are trainable parameters for $k$-hop neighborhood aggregator and $\sigma$ is the activation function implemented as $\text{ReLU}(x) = \max(0, x)$. 
	In Equation~\ref{eq:GCN2}, we assume the neighborhood representation and the target entity representation are integrated via a multi-layer perceptron. After $k$-hop graph convolutional embedding layer, each clicked item is then converted into $\mathbf{i}_t(\mathcal G)= \mathbf{e}_{i_t}^k$.
	
	\subsubsection{Behavior aggregation layer}
	
	Since the interactive recommendation is a sequential decision-making process, at each step, the model requires the current observation of the user as input, and provides a recommended item $i_t$ as output. It is natural to use auto-regressive models such as recurrent neural networks (RNN) to represent the state based on the observation-action sequence ~\cite{hausknecht2015deep,narasimhan2015language}. 
	Thus, we use an RNN with a gated recurrent unit (GRU) as the network cell~\cite{cho2014learning} to aggregate user's historical behaviors and distill user's state $\mathbf{s}_t(\mathcal G)$.
	The update function of a GRU cell is defined as 
	\begin{equation}
	\begin{aligned} 
	\mathbf{z}_{t} &=\sigma_{g}\left(\mathbf{W}_{z} \mathbf{i}_{t}+\mathbf{U}_{z} \mathbf{h}_{t-1}+\mathbf{b}_{z}\right), \\
	\mathbf{r}_{t} &=\sigma_{g}\left(\mathbf{W}_{r} \mathbf{i}_{t}+\mathbf{U}_{r} \mathbf{h}_{t-1}+\mathbf{b}_{r}\right), \\
	\hat{\mathbf{h}}_t &= \sigma_{h}\left(\mathbf{W}_{h} \mathbf{i}_{t}+\mathbf{U}_{h}\left(\mathbf{r}_{t} \circ \mathbf{h}_{t-1}\right)+\mathbf{b}_{h}\right),\\
	\mathbf{h}_{t} &=\left(1-\mathbf{z}_{t}\right) \circ \mathbf{h}_{t-1}+\mathbf{z}_{t} \circ \hat{\mathbf{h}}_t,
	\end{aligned}
	\label{eq:gru}
	\end{equation}
	where $\mathbf{i}_t$ denotes the input vector, $\mathbf{z}_t$ and $\mathbf{r}_t$ denote the update gate and reset gate vector respectively, $\circ$ is the elementwise product operator. 
	The update function of hidden state $\mathbf{h}_t$ is a linear interpolation of previous hidden state $\mathbf{h}_{t-1}$ and a new candidate hidden state $\hat{\mathbf{h}}_t$. The hidden state $\mathbf{h}_t$ is the representation of current user state, which  is then fed into the Q-network, i.e.,
	\begin{equation}
	\mathbf{s}_t(\mathcal G) = \mathbf{h}_t. 
	\label{eq:s_t}
	\end{equation}
	For simplicity, the set of the whole network parameters for computing $\mathbf{s}_t(\mathcal G)$, including parameters of graph convolutional layer and parameters of GRU cell, is denoted as $\theta_S$.

\begin{figure}[t]
		\centering
		\includegraphics[width=\columnwidth]{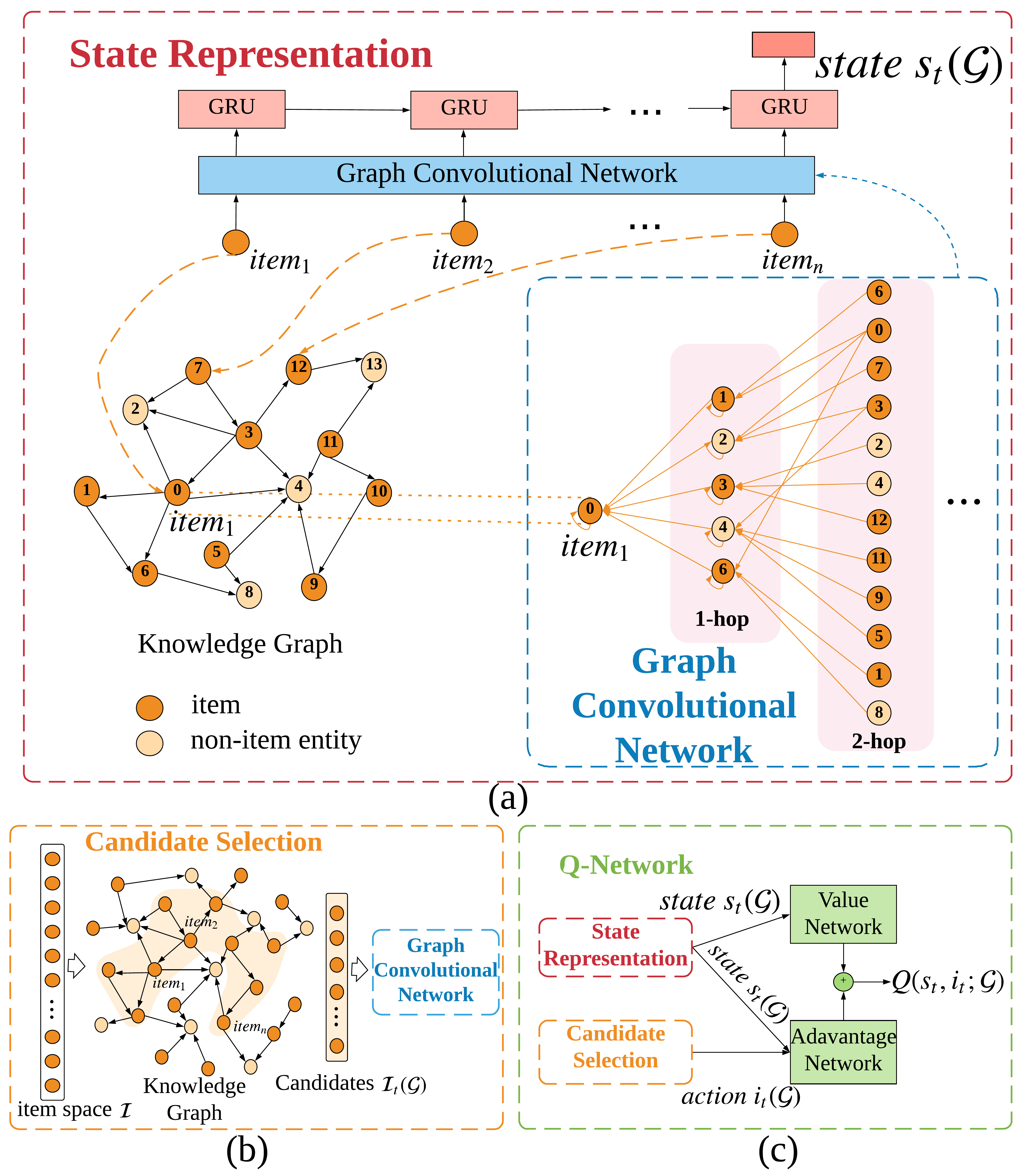}
		\vspace{-10pt}
		\caption{The neural architecture of KGQR. (a) The knowledge-enhanced state representation module maintains user's preferences with a recurrent neural network and a graph neural network; (b) The candidate selection module dynamically reduces large action space according to user's positive feedback; (c) The Q-value network predicts the Q-value with a value network and an advantage network. }
		\label{fig:architecture} 
		\vspace{-10pt}
	\end{figure}

	In Figure~\ref{fig:architecture}(a), we illustrate the knowledge enhanced state representation module elaborated above. The upper part is the recurrent neural network that takes clicked item's embedding at each timestep as the input vector, and outputs the hidden state of the current step as the state representation. The item embeddings, which are the input to GRU, are learned by performing graph convolutional network in KG, as shown in the bottom part.

	\subsection{Neighbor-based Candidate Selection}
	\label{sec:CS}
	Generally, the clicked items have some inherent semantic characteristics, e.g., similar genre movies \cite{wang2018ripplenet}. Since users are usually not likely to be interested in all items, we can focus on selecting the potential candidates for restricted retrieval based on this semantic information in KG. Specifically, we utilize KG to filter some irrelevant items (i.e., actions) and dynamically obtain the potential candidates. The restricted retrieval will focus the data samples on the area that is more useful, as suggested in the structure of item correlation. Thus, these potential candidates would not only reduce the large searching space, but also improve the sample efficiency of policy learning.
	
	More specifically, we perform a sampling strategy based on the $k$-hop neighborhood in KG. 
	In each timestep $t$, the user's historical interacted items serve as the seed set $\mathcal { E } _ { t } ^ {0} = \{i_{1},i_{2},...i_{n}\} $. 
	The $k$-hop neighborhood set starting from the seed entities is denoted as
	\begin{equation}
	\begin{aligned}
	\mathcal { E } _ { t } ^ { k } = \left\{ \textit{tail} ~|~ (\textit{head},\textit{relation}, \textit{tail} ) \in \mathcal { G } \text { and } \textit{head} \in \mathcal { E } _ { t } ^ {l - 1 } \right\} ,\\ \quad l = 1,2 , \ldots ,k .
	\end{aligned}
	\label{eq:CS1}
	\end{equation}
	Then, the candidate action set for the current user state is defined as
	\begin{equation}
	\mathcal{I}_{t}(\mathcal G) =\Big\{ \textit{item} | \textit{item} \in \bigcup_{l=1}^{k}\mathcal { E } _ { t } ^ { l }\  \text{and}\  \textit{item} \in \mathcal{I} \Big\},
	\label{eq:CS2}
	\end{equation}
	with a user-defined cardinality. 
	The shallow part in "candidate selection" in Figure~\ref{fig:architecture}(b) denotes the selected actions with the information from KG. Then all candidate items get their embedding through the graph convolutional layers.

	\subsection{Learning Deep Q-Network} \label{sec:DDQN}

	After modeling the user's state $\mathbf{s}_t(\mathcal G)$ and obtaining candidate sets $\mathcal{I}_{t}(\mathcal G)$, we need to design Q-network to combine this information and improve the recommendation policy for the interactive recommendation process. Here we implement a deep Q-network (DQN) with dueling-Q \cite{wang2015dueling} and double-Q \cite{van2016deep} techniques to model the expected long-term user satisfaction from the current user state as well as to learn the optimal strategy. 
	
	\subsubsection{Deep Q-network.} We adopt dueling technique to reduce the approximation variance and stabilize training ~\cite{wang2015dueling}. That is, using two networks to compute the value function $V(i_{t}(\mathcal G))$ and advantage functions $A(s_{t}(\mathcal G),i_{t}(\mathcal G))$ respectively, and it is shown in Figure~\ref{fig:architecture}. Then the Q-value can be computed as,
	\begin{equation}
	Q(s_t(\mathcal G),i_t(\mathcal G);\theta_V,\theta_A)= V(i_{t}(\mathcal G);\theta_V)+A(s_{t}(\mathcal G),i_{t}(\mathcal G);\theta_A).
	\label{eq:qnet}
	\end{equation}

	Here the approximation of value function and advantage function are accomplished by multi-layer perceptrons. $\theta_V$ and $\theta_A$ is the parameter of value function and advantage function respectively and we denote $\theta_Q = \{\theta_V,\theta_A\}$.
	
	\subsubsection{Model training.} With the proposed framework, we can train the parameters of the model through trial-and-error process. During the interactive recommendation process, at timestep $t$, the recommender agent gets the user's state $\mathbf{s}_t$ from the observations $o_t$ about her, and recommendeds an item $i_t$ via an $\epsilon$-greedy policy (i.e., with probability $1-\epsilon$ choosing the item in the candidate with the max Q-value, with probability $\epsilon$ choosing a random item). Then the agent receives the reward $r_t$ from the user's feedback and stores the experience $(o_t, i_t, r_t, o_{t+1})$ in the replay buffer $\mathcal D$. From $\mathcal D$, we sample mini-batch of experiences, and minimize the mean-square loss function to improve the Q-network, defined as
	\begin{equation}
	L(\theta_Q) = \mathbb{E}_{(o_t,i_t,r_t,o_{t+1})\sim\mathcal D}[(y_t-Q(\mathbf{s}_t,\mathbf{i}_t;\theta_Q))^{2}].
	\label{eq:loss}
	\end{equation}

	Here $y_t$ is the target value based on the optimal $Q^*$. According to Equation (\ref{eq:bellman}), $y_t$ is defined as
	\begin{equation}
	y_t=r_t+\gamma \max\limits_{{i}_{t+1}\in\mathcal I_{t+1}(\mathcal G)}Q(s_{t+1},i_{t+1};\theta_Q).
	\end{equation}
	
	To alleviate the overestimation problem in original DQN, we also utilize a target network $Q'$ along with the online network $Q$ (i.e., the double DQN architecture~\cite{van2016deep}). The online network back-propagates and updates its weights at each training step. The target network is a duplicate of the online network and updates its parameters with training delay. The target value of the online network update is then changed to 	
	\begin{equation}
	y_t=r_t+\gamma Q'\Big(s_{t+1},\mathop{\arg\max}\limits_{i_{t+1}\in\mathcal I_{t+1}(\mathcal G)} Q(s_{t+1},i_{t+1};\theta_Q);\theta_Q'\Big),
	\label{eq:target}
	\end{equation}
	where $\theta_Q'$ denotes the parameter of the target network, and $\theta_Q'$ updates  according to soft assign as
	\begin{equation}
	\theta_Q' = \tau\theta_Q+(1-\tau)\theta_Q',
	\label{eq:Q'}
	\end{equation}
	where the interpolation parameter $\tau$ is also called update frequency. 
	
	To summarize, the training procedure of our KGQR is presented in Algorithm~\ref{alg:trainingKGQR}.
	Note that this paper mainly focuses on the way of incorporating KG into DRL methods for IRS. Thus we study the most typical DQN model as a running example. Our method can be seamlessly incorporated into other DRL models such as policy gradient (PG) \cite{chen2019large}, DDPG \cite{hu2018reinforcement} etc.
	
	\begin{algorithm}[t]
		\small
		\caption{Training KGQR}
		\label{alg:trainingKGQR}
		\KwIn{ $\mathcal{D}$; $\tau$; $\epsilon$;}
		\KwOut{ $\theta_S$;  $\theta_Q$;  $\{\mathbf{e}_h\}_{h\in \mathcal{E}} $; }
		Initialize all parameters: $\theta_S$, $\theta_Q$, $\{\mathbf{e}_h\}_{h\in \mathcal{E}}$, $\theta_Q' \leftarrow \theta_Q$\;
		\Repeat{coverged}{
			\For{$u \in \mathcal U$}{
				Initialize the clicked history  $\mathbf{x} \leftarrow \{\}$\;
				\For{$t \in \{ 0, 1, \ldots , T\}$}{
					for  $ \mathbf{x} =o_t = \{i_{1},i_{2},\ldots,i_{n}\} $, get   $\{ \mathbf{i}_{1}, \ldots, \mathbf{i}_{n}\}$ via Eq.(\ref{eq:GCN1}), Eq.(\ref{eq:GCN2})\;
					Get $\mathbf{s}_t$ via Eq.(\ref{eq:gru}), Eq.(\ref{eq:s_t})\;
					Recommend $i_t$ by $\epsilon$-greedy w.r.t Q-value in Eq.(\ref{eq:qnet})\;
					Receive reward $r_t$\;
					\If{$r_t>0$}{Append $i_t$ to $\mathbf{x}$\;}
					Set $o_{t+1} \leftarrow \mathbf{x}$\;
					Get $\mathcal{I}_{t+1}(\mathcal G)$  via Eq.(\ref{eq:CS1}), Eq.(\ref{eq:CS2})\;
					Store $(o_t, i_t, r_t, o_{t+1}, \mathcal{I}_{t+1}(\mathcal G))$ to buffer $\mathcal{D}$\;
				}
				Sample mini-batch of tuples $(o_t, i_t, r_t, o_{t+1}, \mathcal{I}_{t+1}(\mathcal G)) \sim \mbox{Unif}(\mathcal{D})$\;
				Get  $\mathbf{s}_t,\mathbf{s}_{t+1}$ from $o_{t}, o_{t+1}$  via Eq.(\ref{eq:GCN1}), Eq.(\ref{eq:GCN2}) and Eq.(\ref{eq:gru})\;
				Construct target values $y_t$ via Eq. (\ref{eq:target})\;
				Update $\theta_S$, $\theta_Q$, $\{\mathbf{e}_h\}_{h\in \mathcal{E}} $ via  SGD w.r.t the loss function Eq.(\ref{eq:loss})\;
				Update $\theta_Q'$ via Eq. (\ref{eq:Q'}) \;
			}
		}
	\end{algorithm}

	\section{Experiment}
	\label{sec:exp}
	We conduct experiments on two real-world datasets to evaluate our proposed KGQR framework. 
	We aim to study the following research questions (RQs):
	\begin{itemize}[leftmargin = 10pt]
		\item \textbf{RQ1:} How does KGQR perform as compared with state-of-the-art interactive recommendation methods?
		\item \textbf{RQ2:} Does KGQR improve sample efficiency?
		\item \textbf{RQ3:} How do different components (i.e., KG-enhanced state representation, GCN-based task-specific representation learning, neighbor-based candidate selection) affect the performance of KGQR? 
	\end{itemize}
	
	\subsection{Experimental Settings}
	\subsubsection{Datasets}
	We adopt two real-world benchmark datasets for evaluation and describe them as below.
	\begin{savenotes}
		\begin{description}[leftmargin = 10pt]
			\item[Book-Crossing\footnote{http://www2.informatik.uni-freiburg.de/$\sim$\text{cziegler/BX/}}] is a book rating dataset from Book-Crossing community. The ratings are ranging from 0 to 10. This dataset is linked with Microsoft Satori and the sub-KG is released by \cite{wang2018ripplenet}.
			\item[Movielens-20M\footnote{https://grouplens.org/datasets/movielens/}] is a benchmark dataset, which consists of 20 million ratings from users to movies in MovieLens website. The ratings are ranging from 1 to 5. It is also linked with Microsoft Satori and the sub-KG is released by \cite{wang2019knowledge}.
		\end{description}
	\end{savenotes}
	For Book-Crossing dataset, we follow the processing of \cite{wang2018ripplenet} to convert original ratings into two categories, 1 for high ratings, 0 for others. For MovieLens-20M dataset, we keep the users with at least 200 interactions. The statistics information of these two datasets is presented in Table~\ref{tab:dataset}.
	
	We choose these two typical datasets since our work focuses on incorporating KG into RL-based models for IRS. The experiments on more datasets with rich domain information such as news or images will be left as future work.

	\begin{table}[]
		\centering
		\caption{Statistics of the datasets.}
		\vspace{-10pt}
		\small
		\begin{tabular}{cl|r|r}
			\hline
			& &Book-crossing & Movielens-20M \\
			\hline
			\hline
			\multirow{3}{*}{\begin{tabular}[c]{@{}c@{}}User-Item\\ Interaction\end{tabular}} & \#User &17,860 & 16,525\\
			&\#Linked Items &14,910&16,426\\
			&\#Interactions &139,746&6,711,013 \\
			\hline
			\hline
			\multirow{3}{*}{\begin{tabular}[c]{@{}c@{}}Knowledge\\ Graph\end{tabular}} & \#Entities &77,903 &101,769 \\
			&\#Relation Types &25&32\\
			&\#Triples &151,500& 489,758\\
			\hline   
		\end{tabular}
		\label{tab:dataset}
	\end{table}
	
	\subsubsection{Simulator}
	Due to the interactive nature of our problem, online experiments where the recommender system interacts with users and learns the policy according to the users' feedback directly would be ideal. However, as mentioned in Section~\ref{sec:intro}, the trial-and-error strategy for training policy in an online fashion would degrade the user's experience, as well as the system profit. Thus, the community has formed a protocol \cite{hu2018reinforcement,chen2019large,zhao2018deep,dulac2015deep,wang2017factorization} to build up an environment simulator based on offline datasets for evaluation. 
	
	Following the experiment protocol in \cite{chen2019large}, our mimic environment simulator takes into account the instinctive feedback as well as the sequence nature of user behavior. We perform matrix factorization to train the 20-dimensional embeddings of the users and items. Then we normalize the ratings of each dataset into the range [-1,1], and use them as users' instinctive feedback. Then we combine a sequential reward with the instinctive reward. For instance, if the recommender system recommends an item $i_{j}$ to a user $u_{i}$ at timestep $t$, the final reward function comes as
	\begin{equation}\label{eq:sim-reward}
	R(s_{t},i_{t}) = r_{ij} + \eta(c_{p}-c_{n}),
	\end{equation}
	where $r_{ij}$ is the predicted rating given by the simulator, $c_{p}$ and $c_{n}$ means the consecutive positive and negative counts representing the sequential pattern, and $\eta$ is a trade-off between instinctive feedback and sequential nature. In our experiment, $\eta$ is chosen from \{0.0, 0.1, 0.2\}, following the empirical experience in~\cite{chen2019large}. 
	
	For each dataset, we randomly divide the users into two parts: 80\% of the users are used for training the parameters of our model, and the other 20\% are used for testing the model performance. 
	Due to train/test dataset splitting style, the users in our test set never exist in the training set. 
	That is to say, the experiment is a cold-start scenario, which means there is no user click history at the beginning. 
	To handle cold-start problem, the recommender system collects the most popular items among the training users, and recommends a popular item to a test user at step $t_0$. 
	Then, according to the user's feedback, the recommender system recommends items to the user interactively. Besides, we remove the recommended items from the candidate set to avoid repeated recommendation in one episode. 
	The episode length $T=32$ for all the two datasets in our experiment.

	\subsubsection{Evaluation Metrics}
	Three evaluation metrics are used.\\
	\noindent\textbf{Average Reward}. As an IRS aims to maximize the reward of the whole episode, a straightforward evaluation measure is the average reward over each interaction of test users.
	\begin{equation}
	Reward =  \frac{1}{\#users \times T}\sum_{users}\sum_{t=1}^{T}\gamma ^{t}R(s_{t},i_{t})
	\end{equation}
	We also check for the precision and recall during $T$ timesteps of the interactions, which are widely used metrics in traditional recommendation tasks.\\
	\noindent\textbf{Average Cumulative Precision@$T$}.
	\begin{equation}
	Precision@T = \frac{1}{\#users\times T}\sum_{users}\sum_{t=1}^{T}\theta_{hit}
	\end{equation}
	\noindent\textbf{Average Cumulative Recall@$T$}.
	\begin{equation}
	Recall@T = \frac{1}{\#users}\sum_{users}\sum_{t=1}^{T}\frac{\theta_{hit}}{\#preferences}
	\end{equation}
	We define $\theta_{hit}=1$ if the instinctive feedback of the recommended item given by the simulator is higher than the predefined threshold, which is 0.5 in Book-Crossing and 3.5 in Movielens-20M. And we define $\#preferences$ is the total number of the positive instinctive feedback among all items, i.e., number of items with $r_{ij}>\textit{boundary}$ based on the simulator.
	
	\noindent\textbf{Significance test}. The Wilocoxon signed-rank test has been performed to evaluate whether the difference between KGQR and the other baselines is significant.
	
	\subsubsection{Baselines}
	
	We compare KGQR with 7 representative baseline methods in the IRS scenario, where GreedySVD and GRU4Rec are traditional recommendation methods, LinearUCB and HLinearUCB are based on multi-armed bandits, DDPG$^\text{KNN}$, DDPGR, DQNR are DRL-based methods. 
	
	\begin{description}[topsep = 3pt,leftmargin =10pt]
		\item [GreedySVD] is a well-known collaborative filtering methods via singular value decomposition \cite{koren2008factorization}. In interactive scenarios, we train the model after each interaction with users and recommend an item with the predicted highest rating to one user.
		\item [GRU4Rec] is a representative RNN-based sequential recommendation algorithm~\cite{hidasi2015session} to predict what the user will click at the next timestep based on the browsing histories. 
		\item[LinearUCB] is a multi-armed bandit algorithm~\cite{li2010contextual} which selects items according to the estimated upper confidence bound (UCB) of the potential reward based on contextual information about the users and items.
		\item[HLinearUCB] is a contextual bandit algorithm combined with extra hidden features~\cite{wang2017factorization}.
		\item[DDPG$^\text{KNN}$] is a DDPG-based method~\cite{dulac2015deep} which represents the discrete actions into a continuous vector space. The actor selects a proto-action in the continuous space and then chooses the item with the max Q-value from the candidate items selected via $K$-nearest-neighbor (KNN) according to the proto-action. In this approach, a larger $K$ value boosts the performance but also brings computational cost, indicating the existence of a trade-off between performance and efficiency. In our experiment, the $K$ value is set to \{1, 0.1$N$, $N$\}, where $N$ is the total number of items.
		\item[DDPGR] is a DDPG-based method~\cite{zhao2018deep} where the actor learns a ranking vector. The vector is utilized to compute the ranking score of each item, by performing product operation between this vector and item embedding. Then the item with the highest ranking score is recommended to the user.
		\item[DQNR] is a DQN-based method~\cite{zheng2018drn} where the recommender system learns a Q function to estimate the Q-value of all the actions at a given state. The method then recommends the item with highest Q-value at the current state. 
	\end{description}
	
	Note that, traditional knowledge enhanced recommendation methods like CKE~\cite{zhang2016collaborative}, RippleNet~\cite{wang2018ripplenet}, KGAT~\cite{wang2019kgat} and etc. are not suitable for the online interactive recommendation as tested in this paper. Because our recommendation process is an online sequential recommendation in the case of cold-start setting, this means there is no data about the test user at the beginning. We model user preferences in real-time through the user interaction process and provide recommendations under the current situation. These traditional models could not handle this cold-start problem; thus, we do not compare our proposed model with them.
	
	\subsubsection{Parameter Settings}
	In KGQR, we set the maximal hop number $k=2$ for both datasets. We have tried larger hops, and find that the model with larger hops brings exponential growth of computational cost with only limited performance improvement. 
	The dimension of item embedding is fixed to 50 for all the models. For baseline methods, the item embedding is pre-trained by matrix factorization with the training set. For KGQR, we pre-train the embedding of KG by TransE~\cite{bordes2013translating}, and then embedding will be updated while learning the deep Q-network. Besides, other parameters are randomly initialized with uniform distribution. The policy network in all the RL-based methods takes two fully-connected layers with activation function as ReLU. The hyper-parameters of all models are chosen by grid search, including learning rate, $L_{2}$ norm regularization, discount factor $\gamma$ and etc. All the trainable parameters are optimized by \emph{Adam} optimizer ~\cite{kingma2014adam} in an end-to-end manner. We use PyTorch \cite{paszke2019pytorch} to implement the pipelines and train networks with an NVIDIA GTX 1080Ti GPU. 
	We repeat the experiments 5 times by changing the random seed for KGQR and all the baselines. 

	\begin{table*}[]
		\centering
		\caption{Overall Performance Comparison.}
		\vspace{-10pt}
		\resizebox{\textwidth}{!}{
			\begin{tabular}{c|c|c|c|c|c|c|c|c|c|c}
				\hline
				\multirow{2}{*}{Dataset} & \multirow{2}{*}{Method} & \multicolumn{3}{c|}{$\eta=0$}           & \multicolumn{3}{c|}{$\eta=0.1$}                  & \multicolumn{3}{c}{$\eta=0.2$} \\ \cline{3-11} 
				&                         & Reward & Precision@32 & Recall@32  & Reward & Precision@32 & Recall@32  & Reward & Precision@32 & Recall@32 \\ \hline \hline
				
				\multirow{8}{*}{Book-Crossing}& Greedy SVD & -0.0890 & 0.3947 & 0.0031 &  -0.1637 & 0.4052 & 0.0032 & -0.2268 & 0.4133 & 0.0033  \\ 
				& GRU4Rec & 0.5162 & 0.8611 &0.0070  & 1.3427  & 0.8595  & 0.0070 &2.1797 & 0.8625  & 0.0070 \\  \cline{2-11}
				&  LinearUCB & -0.0885 & 0.3956 & 0.0032 & -0.1640 & 0.4049 & 0.0032  & -0.2268 & 0.4133 & 0.0033  \\
				& HLinearUCB& -0.1346 & 0.3819 & 0.0031  & -0.3566 & 0.3841 & 0.0031  & -0.6064 & 0.3713 & 0.0031 \\ \cline{2-11} 
				& DDPGR & 0.5521 & 0.9115 & 0.0074  & 1.1412 & 0.8800 & 0.0072  & 2.2057 & 0.9270 & 0.0076  \\				
				& DDPG$^\text{KNN}$(K=1) & 0.3159 & 0.7302 & 0.0059  & 0.7312 & 0.7990 & 0.0065  & 0.8409 & 0.7472 & 0.0061  \\
				& DDPG$^\text{KNN}$(K=0.1N) & 0.7312 & 0.9907 & 0.0080  & 2.0750 & 0.9813 & 0.0080  & 3.3288 & 0.9758 & 0.0079 \\
				& DDPG$^\text{KNN}$(K=N) & 0.7639 & 0.9927 & 0.0081  & 2.2729 & 0.9942 & 0.0081 & 3.7179 & 0.9915 & 0.0081  \\
				& DQNR & 0.7634 & 0.9936 & 0.0081 & 2.2262 & 0.9907 & 0.0080 & 3.6118 & 0.9881 & 0.0080 \\  \cline{2-11}
				& KGQR &\textbf {0.8307*} & \textbf{0.9945*} & \textbf{0.0081}  & \textbf{2.3451*}& \textbf{0.9971*} &\textbf{0.0081}  & \textbf{3.7661*}& \textbf{0.9966*} &\textbf{0.0081*} \\ \hline\hline
				
				\multirow{8}{*}{MovieLens-20M} & Greedy SVD &0.4320  &0.6569  & 0.0049&  0.6915 &0.6199  & 0.0048    &0.9042 & 0.5932 &  0.0046\\ 
				& GRU4Rec &  0.7822 & 0.8382 & 0.0074 &  1.5267 & 0.8253  & 0.0072  & 2.3500 & 0.8316  & 0.0073   \\  \cline{2-11}
				& LinearUCB & 0.2307  &0.3790  &0.0029   &0.6147 & 0.5821 & 0.0046  &  0.8017  &0.5614  &0.0044    \\
				& HLinearUCB & 0.0995 & 0.3852 & 0.0029  & 0.0172 &0.3841  &  0.0028& 0.2265 & 0.3774 & 0.0027  \\  \cline{2-11}
				& DDPGR &0.2979 &0.4917 &0.0034   &1.4952  & 0.7626 &  0.0055 &2.3003  & 0.6977 &0.0045    \\				
				& DDPG$^\text{KNN}$(K=1) &0.5755  &  0.7293&  0.0059& 1.0854 & 0.7165 &0.0058   & 1.6912 & 0.7371 & 0.0061 \\
				& DDPG$^\text{KNN}$(K=0.1N) & 0.6694 & 0.8167 &0.0070   & 1.1578 & 0.8165 & 0.0069 &2.2212  & 0.8215 & 0.0068  \\
				& DDPG$^\text{KNN}$(K=N) & 0.8071 & 0.9606&  0.0082 &2.1544 & 0.9446 & 0.0081  & 3.6071 & 0.9533 & 0.0082 \\
				& DQNR &0.8863 & 0.9680 &0.0086    &2.3025  & 0.9667 & 0.0081   &3.4036  & 0.9089 &  0.0071 \\  \cline{2-11}
				& KGQR & \textbf{0.9213*}  &\textbf{0.9726*} & \textbf{0.0086*} &  \textbf{2.4242*}& \textbf{0.9722*} & \textbf{0.0083*}&\textbf{3.7695*} &  \textbf{0.9713*}&\textbf{0.0084*} \\ \hline
			\end{tabular}
		}
		\footnotesize \flushleft{* indicates statistically significant improvements (measured by Wilocoxon signed-rank test at $p<0.05$) over all baselines.}
		\label{results}
	\end{table*}

	\subsection{Overall Performance (RQ1)}
	Table~\ref{results} reports the performance comparison results. We have the following observations:
	(i) KGQR consistently obtains the best performance across all environment settings on both datasets. For instance, compared to RL-based recommendation methods like DQN-based (i.e., DQNR) and DDPG-based (e.g., DDPG$^\text{KNN}$, DDPGR), KGQR improves over the strongest baselines in terms of Reward by 3.2\% and 5.3\% in Book-Crossing and MovieLens-20M, respectively. For traditional evaluation metrics, KGQR improves Precision@32 by 0.5\% and 1.9\% in the two datasets, respectively. This demonstrates that the leverage of prior knowledge in KG significantly improves the recommendation performance.
	(ii) In most conditions, non-RL methods including conventional methods and MAB-based methods, perform worse than the RL-based methods. Two reasons stand for the significant performance gap. On the one hand, except GRU4Rec, the capacity of other non-RL methods are limited in modeling user preference without considering sequential information. On the other hand, they all focus on the immediate item reward and do not take the present value of the overall performance of the whole sequence into the current decision, which makes them perform even worse in environments that give future rewards more (e.g., $\eta=0.1$, $\eta=0.2$).
	(iii) Among the RL-based baselines, we can observe that DQNR and DDPG$^\text{KNN}$ ($K=N$) achieves much better performance than the other DDPG based methods. 
	When $K=N$ ($N$ is the total number of items),  DDPG$^\text{KNN}$ can be seen as a greedy policy that always picks the item with max Q-value. 
	We also notice that the training process of DDPG based methods is not stable, e.g., their training curves sometimes experience a sudden drop. This may be accounted for that the continuous proto-action picked by the actor is inconsistent with the final action that the critic is learned with.
	Such inconsistency between actor and critic may result in inferior performance.

	\begin{figure}[t]
		\centering
		\includegraphics[width=0.5\textwidth]{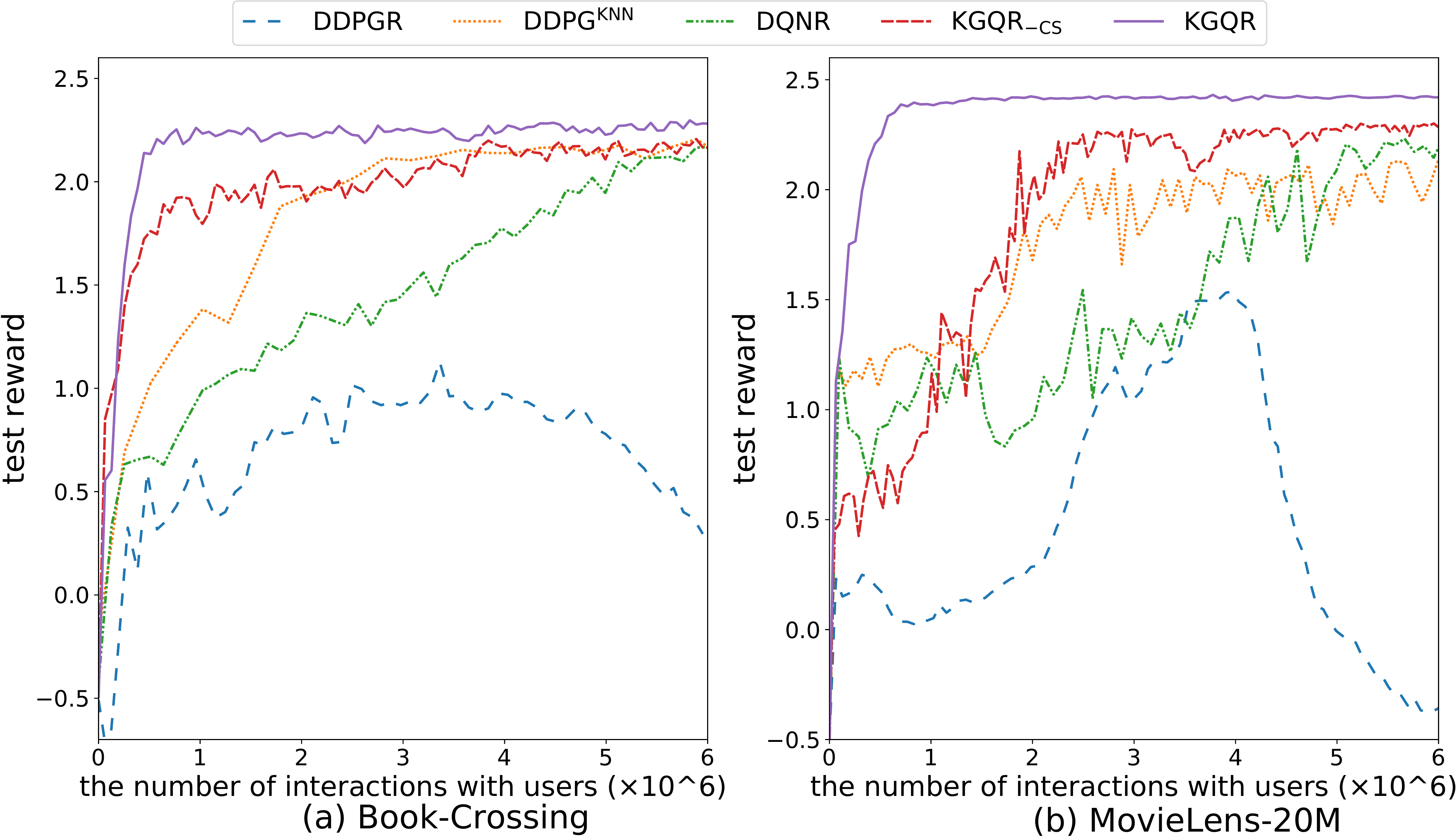}
		\vspace{-10pt}
		\caption{Learning curves of KGQR and DRL-based baseline models. }
		\label{fig:sample efficiency} 
		\vspace{-10pt}
	\end{figure}
	
	\begin{table}[]
		\centering
		\caption{Sample Efficiency Comparison: number of interactions to achieve reward 0.5,1.0,1.5,2.0 for each dataset.}
	    \vspace{-10pt}
		\resizebox{0.50\textwidth}{!}{
			\begin{tabular}{c|cccc|cccc}
				\hline
				\multirow{2}{*}{Model} & \multicolumn{4}{c|}{Book-Crossing}  &\multicolumn{4}{c}{Movielens-20M} \\ \cline{2-9} 
				&0.5& 1.0 & 1.5&2.0  & 0.5& 1.0&1.5&2.0  \\
				\hline
				\hline
				DDPGR  &0.46M &2.49M &- & -&2.28M &2.62M &3.82M &- \\
				DDPG$^\text{KNN}$ &0.20M &0.49M & 1.46M& 2.44M&0.04M &0.07M&1.76M&2.42M\\
				DQNR&0.20M &1.06M & 3.09M&4.83M &0.06M &0.08M&2.47M&4.27M\\
				KGQR &0.06M &0.17M &0.24M &0.40M & 0.04M&0.06M&0.16M&0.33M\\
				\hline 
			\end{tabular}
		}
		\label{tab:sample efficiency}
		
	\end{table}

	\subsection{Sample Efficiency (RQ2)}

	One motivation of exploiting KG is to improve sample efficiency in RL-based recommendation, i.e., to reduce the amount of interaction data needed to achieve the same performance. In Figure~\ref{fig:sample efficiency} and Table~\ref{tab:sample efficiency}, we analyze the number of interactions needed for each DRL-based model to achieve the same performance with the environment of $\eta=0.1$ in Eq.(\ref{eq:sim-reward}). As can be observed, our proposed KGQR can achieve the same performance as the other RL-based methods using the least number of interactions. More specifically, to achieve a test reward of 2.0, KGQR only needs 17.3\% and 13.6\% of interactions compared to the second efficient baseline (i.e., DDPG$^\text{KNN}$) in the two datasets. This result empirically validates that sample efficiency is improved by utilizing the semantic and correlation information of items in KG. The detailed analysis of different components that contributes to improve sample efficiency is proposed in Section~\ref{sec:sample}.

	\begin{table}[]
		\centering
		\caption{Comparison of Different KGQR Variants.}
		\vspace{-10pt}
		\begin{tabular}{c|c|c|c|c}\hline
			& KGQR\footnotesize-KG\normalsize & KGQR\footnotesize-GCN-CS\normalsize & KGQR\footnotesize-CS\normalsize  & KGQR   \\\hline\hline
			KGemb*   & $\times$ & $\checkmark$ &$\checkmark$ & $\checkmark$\\
			GCNprop* & $\times$  & $\times$  & $\checkmark$ & $\checkmark$\\
			CS*  & $\times$   &  $\times$  & $\times$ & $\checkmark$\\
			\hline   
		\end{tabular}
		\label{tab:KGQR-variants}
		\footnotesize \flushleft{* KGemb denotes KG enhanced item representation; GCNprop denotes GCN propagation in state representation; CS denotes the neighbor-based candidate selection.}
	\end{table}
	
	\begin{table}[]
		\centering
		\caption{Ablation Study of KGQR.}
		\vspace{-10pt}
		\resizebox{0.48\textwidth}{!}{
			\begin{tabular}{c|cc|cc}
				\hline
				\multirow{2}{*}{Model} & \multicolumn{2}{c|}{Book-Crossing}  &\multicolumn{2}{c}{Movielens-20M} \\ \cline{2-5} 
				& Reward & Precision@32  & Reward & Precision@32  \\
				
				\hline
				\hline
				KGQR\footnotesize-KG\normalsize   &2.2262 & 0.9907&2.3025 & 0.9667  \\
				KGQR\footnotesize-GCN-CS\normalsize &2.2181&0.9819& 2.2402 & 0.9621\\
				KGQR\footnotesize-CS\normalsize&  2.2836  &0.9939   & 2.3689&0.9698\\
				\textbf{KGQR}  & \textbf{2.3451} & \textbf{0.9971}   & \textbf{2.4242} & \textbf{0.9722} \\
				\hline 
		\end{tabular}}
		\label{tab:ablation study}
	\end{table}
	
	\subsection{Analysis (RQ3)}
	
	In this section, we further analyze the effectiveness of different components in the proposed framework. In KGQR, there are three components utilizing KG that may affect the performance of KGQR: KG enhanced item representation, GCN propagation in state representation (Section~\ref{sec:state}) and neighbor-based candidate selection (Section~\ref{sec:CS}). To study the effectiveness of each such component, we evaluate the performance of four different KGQR variants, namely KGQR\footnotesize-KG\normalsize~(i.e.,DQNR), KGQR\footnotesize-CS\normalsize, KGQR\footnotesize-GCN-CS\normalsize~and KGQR. The relationship between KGQR variants and different components is presented in Table~\ref{tab:KGQR-variants}. In the ablation study, we consider the environment of $\eta=0.1$ in Eq.(\ref{eq:sim-reward}), and Table~\ref{tab:ablation study} shows the performance of these four variants.

	\begin{figure}[t]
		\centering
		\begin{subfigure}[b]{0.45\columnwidth}
			\includegraphics[width=\textwidth]{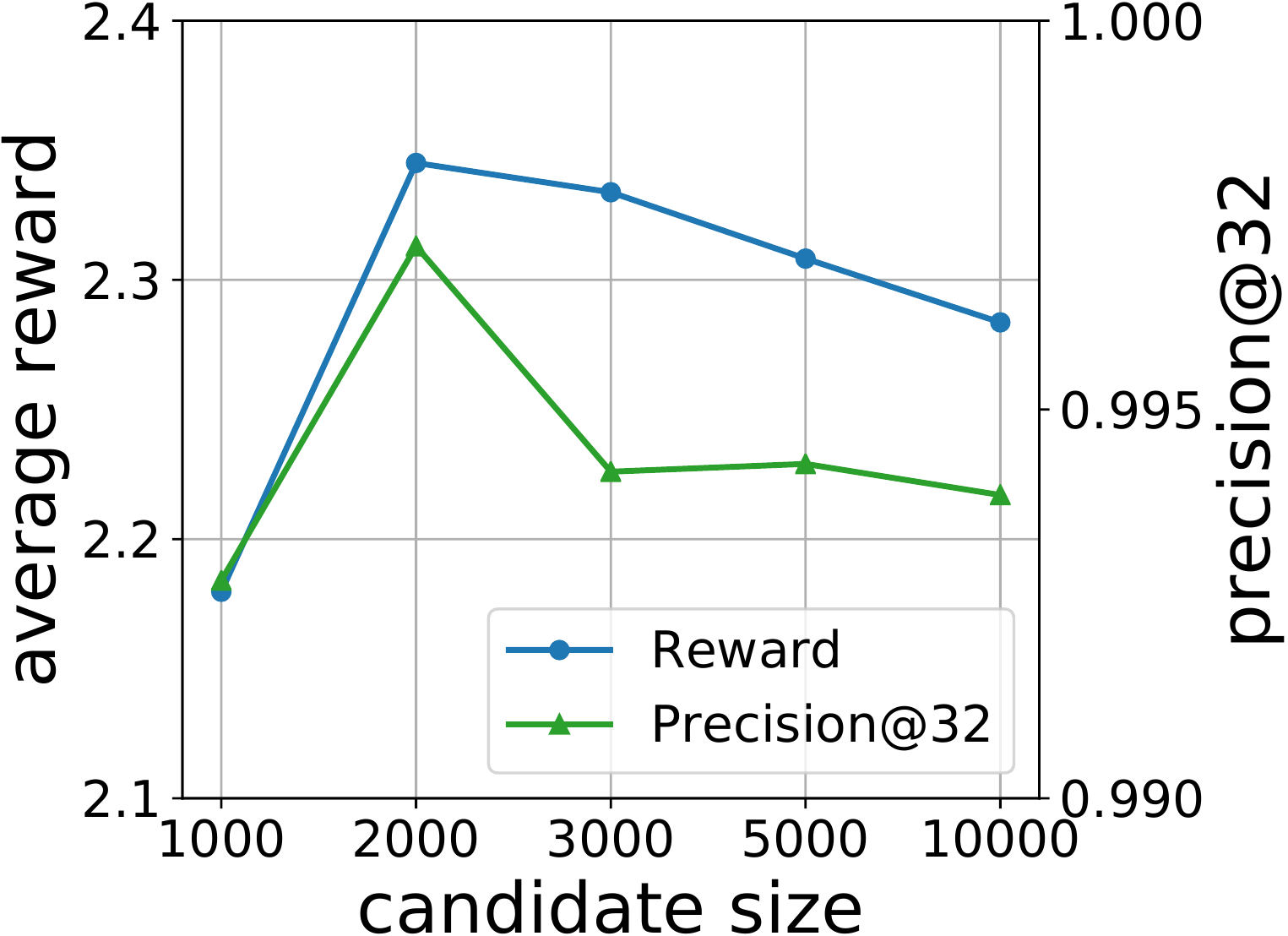}
			\caption{Book-Crossing}
		\end{subfigure}
		\hfill
		\begin{subfigure}[b]{0.45\columnwidth}
			\includegraphics[width=\textwidth]{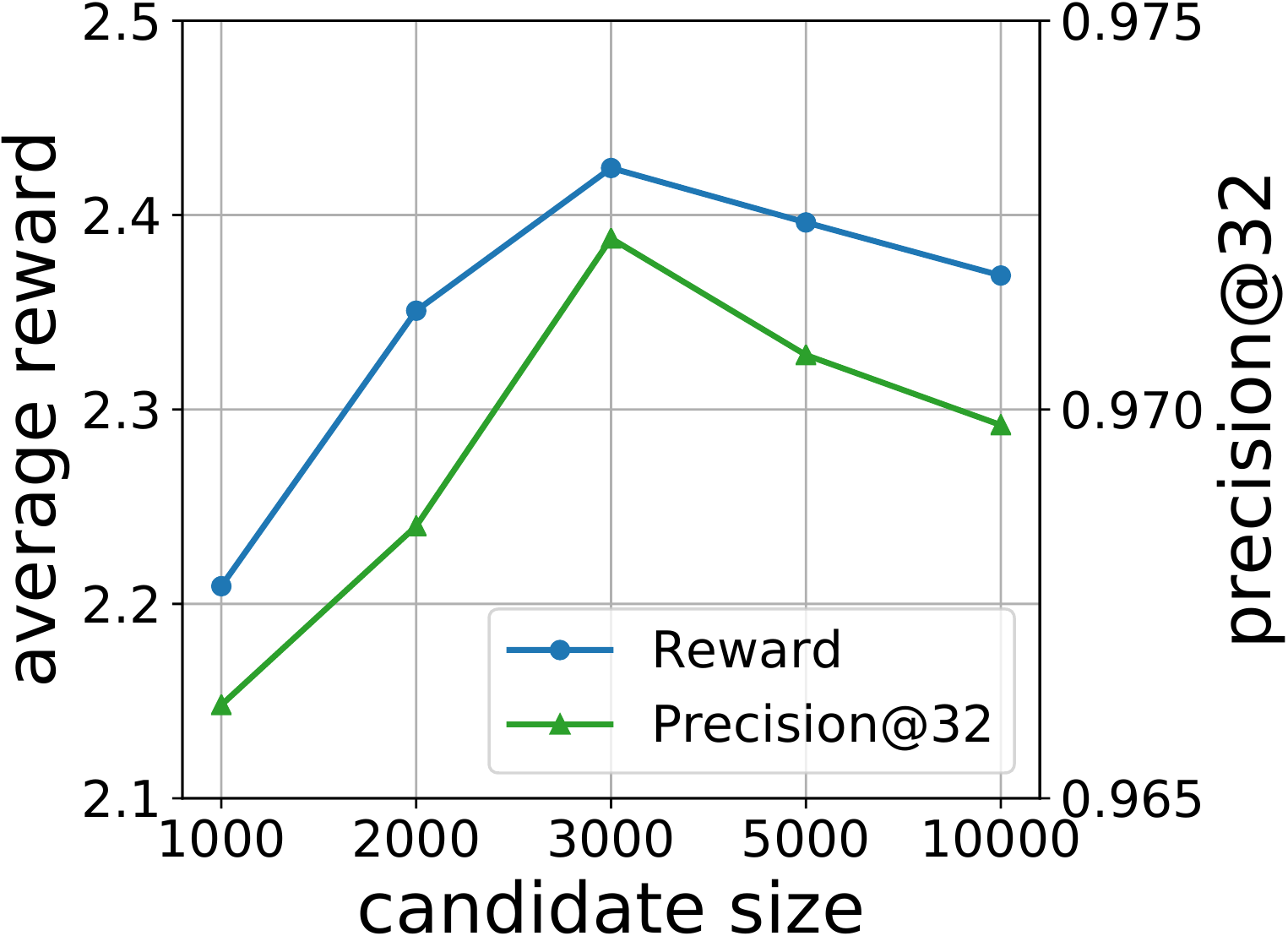}
			\caption{MovieLens-20M}
		\end{subfigure}
		\vspace{-10pt}
		\caption{Influence of Candidate Size. }
		\label{fig:candidate selection} 
	\end{figure}

	\subsubsection{Recommendation performance}
	\noindent \textbf{Effect of KG enhanced item representation.}
	In KGQR\footnotesize-KG\normalsize, the item embeddings are pretrained by MF model from the historical interaction data, while in KGQR\footnotesize-GCN-CS\normalsize, they are retrieved from KG, pretrained with TransE. Therefore, the marginal difference between KGQR\footnotesize-KG\normalsize~and KGQR\footnotesize-GCN-CS\normalsize~performance indicates that the information in KG has almost equal contribution with the historical interaction data, which suggests the superiority of applying KG for cold-start scenario (i.e., no historical interaction data exists). 
	
	\noindent \textbf{Effect of GCN propagation in state representation.} Comparing the performance of KGQR\footnotesize-GCN-CS\normalsize~with KGQR\footnotesize-CS\normalsize~in Table~\ref{tab:ablation study}, the improvement in KGQR\footnotesize-GCN-CS\normalsize~ indicates that the signal from RL-based recommendation guides the update of KG embedding so that the items in KG can be represented more suitably for the current specific recommendation task.
	
	\noindent \textbf{Effect of neighbor-based candidate selection.} The comparison between the performance of KGQR\footnotesize-CS\normalsize~and KGQR validates the effectiveness of neighbor-based candidate selection, for candidate selection module can leverage the local structure of interacted items in KG to filter irrelevant items and such restricted retrieval improves the final recommendation performance. 
	
	To study the influence of candidate size, we vary the candidate size in the range of \{1000, 2000, 3000, 5000, 10000\} and present the recommendation performance in Figure~\ref{fig:candidate selection}. We observe that recommendation performance first grows as candidate size increases, since a small size of candidate limits possible choices of the recommendation algorithm. However, further increasing of candidate size degrades the performance, since the neighbor-based candidate selection filters some irrelevant items in advance. Such irrelevant items have very limited chance to be recommended and to collect feedback which makes them not be able to learn well by the recommendation algorithm and eventually gains a negative effect to the performance.

	\subsubsection{Sample efficiency}
	\label{sec:sample}
	
	\noindent  \textbf{Effect of KG-enhanced state representation.} Comparing the number of interactions of KGQR\footnotesize-KG\normalsize~ with that of KGQR w.r.t same test average reward in Figure~\ref{fig:sample efficiency}, we notice that in both two environments the utilizing of KG and task-specific representation learning improves the sample efficiency. This observation demonstrates our motivation that the propagation of user preference through the correlated items via GCN is helpful in dealing with sample efficiency issues. 
	
	\noindent \textbf{Effect of neighbor-based candidate selection.} Besides the performance improvements, the candidate selection significantly improves the sample efficiency, as shown in Figure~\ref{fig:sample efficiency} (comparing the purple line and red line).

	\section{Conclusion}
	\label{sec:conclude}
	In this work, we proposed a knowledge graph enhanced Q-learning framework (KGQR) for the interactive recommendation. To the best of our knowledge, it is the first work leveraging KG in RL-based interactive recommender systems, which to a large extent addresses the sample complexity issue and significantly improves the performance. 
	Moreover, we directly narrow down the action space by utilizing the structure information of knowledge graphs to effectively address the large action space issue. 
	The model propagates user preference among the correlated items in the graph, to deal with the extremely sparse user feedback problem in IRS.
	All these designs improve sample efficiency, which is a common issue in previous works.
	The comprehensive experiments with a carefully-designed simulation environment based on two real-world datasets demonstrate that our model can lead to significantly better performance with higher sample efficiency compared to state-of-the-arts.
	
	For future work, we plan to investigate KGQR on news and image recommendation tasks with other DRL frameworks such as PG and DDPG. We are also scheduling the process of deploying KGQR onto an online commercial recommender system. Further, we are interested in inducing a more complex sequential model to represent the dynamics of user preferences, e.g., taking user's propensity to different relations that the click history shows into consideration.

	\section*{Acknowledgement}
    The corresponding author Weinan Zhang thanks the support of "New Generation of AI 2030" Major Project 2018AAA0100900 and NSFC (61702327, 61772333, 61632017, 81771937). The work is also sponsored by Huawei Innovation Research Program.

	\bibliographystyle{ACM-Reference-Format}
	\bibliography{sample-base}

\end{document}